\documentclass[conference]{IEEEtran}

\usepackage{graphicx}
\usepackage{epstopdf}
\usepackage{standalone}
\usepackage[nolist ]{acronym}
\usepackage{tcolorbox}

\usepackage{pdfpages}
\usepackage{tikz}
\usetikzlibrary{positioning}
\usepackage[bookmarks=false]{hyperref}
\usepackage{pdfcomment}
\usepackage{hologo}

\usepackage{xstring}

\usepackage[utf8]{inputenc}
\usepackage{marvosym}

\usepackage{algorithm}
\usepackage{algpseudocode}
\usepackage{tikz}

\usepackage{listings}
\definecolor{ListingBackground}{rgb}{0.97,0.97,0.97}
\usepackage{pgfplots}
\pgfplotsset{compat=newest}
\pgfplotsset{
	box plot/.style={
		/pgfplots/.cd,
		fill=blue!30,
		only marks,
		mark=-,
		mark size=0.2em,
		/pgfplots/error bars/.cd,
		y dir=plus,
		y explicit,
	},
	box plot box/.style={
		/pgfplots/error bars/draw error bar/.code 2 args={%
			\draw  ##1 -- ++(.2em,0pt) |- ##2 -- ++(-.2em,0pt) |- ##1 -- cycle;
		},
		/pgfplots/table/.cd,
		y index=2,
		y error expr={\thisrowno{3}-\thisrowno{2}},
		/pgfplots/box plot
	},
	box plot top whisker/.style={
		/pgfplots/error bars/draw error bar/.code 2 args={%
			\pgfkeysgetvalue{/pgfplots/error bars/error mark}%
			{\pgfplotserrorbarsmark}%
			\pgfkeysgetvalue{/pgfplots/error bars/error mark options}%
			{\pgfplotserrorbarsmarkopts}%
			\path ##1 -- ##2;
		},
		/pgfplots/table/.cd,
		y index=4,
		y error expr={\thisrowno{2}-\thisrowno{4}},
		/pgfplots/box plot
	},
	box plot bottom whisker/.style={
		/pgfplots/error bars/draw error bar/.code 2 args={%
			\pgfkeysgetvalue{/pgfplots/error bars/error mark}%
			{\pgfplotserrorbarsmark}%
			\pgfkeysgetvalue{/pgfplots/error bars/error mark options}%
			{\pgfplotserrorbarsmarkopts}%
			\path ##1 -- ##2;
		},
		/pgfplots/table/.cd,
		y index=5,
		y error expr={\thisrowno{3}-\thisrowno{5}},
		/pgfplots/box plot
	},
	box plot median/.style={
		/pgfplots/box plot
	},
	boxplot/every median/.style={
		ultra thick,dashed,cyan
	}
}

\definecolor{flexicolor}{RGB}{46,49,146}
\definecolor{amaricolor}{RGB}{237,28,36}

\usepackage{xspace}

\usepackage[binary-units=true]{siunitx}
\usepackage{psfrag}
\usepackage{graphicx}
\usepackage{tabularx,booktabs}
\usepackage{multirow}
\usepackage{rotating}

\renewcommand{\baselinestretch}{1}

\usepackage{multirow}

\usepackage[cmex10]{amsmath}
\usepackage[caption=false,font=footnotesize]{subfig}
\usepackage{stfloats}
\begin{document}

\newcommand{\paperTitle}{Performance Evaluation and Optimization of B.A.T.M.A.N. V Routing for Aerial and Ground-based Mobile Ad-hoc Networks}
\newcommand{\paperAuthors}{Benjamin Sliwa, Stefan Falten and Christian Wietfeld}
\newcommand{\paperEmails}{$\{$Benjamin.Sliwa, Stefan.Falten, Christian.Wietfeld$\}$@tu-dortmund.de}
\newcommand{\githubUrl}{\footnote{Available at \colorbox{red}{TODO GITHUB URL}}\xspace}

\newcommand{\figurePadding}{0pt}
\newcommand{\figureTopPadding}{\figurePadding}
\newcommand{\figureBottomPadding}{\figurePadding}

\newcommand\single{1\textwidth}
\newcommand\double{.48\textwidth}
\newcommand\triple{.32\textwidth}
\newcommand\quarter{.24\textwidth}
\newcommand\singleC{1\columnwidth}
\newcommand\doubleC{.475\columnwidth}

\renewcommand{\vec}[1]{\mathbf{#1}}

\newcommand{\dummy}[3]
{
	\begin{figure}[b!]  
		\begin{tikzpicture}
		\node[draw,minimum height=6cm,minimum width=\columnwidth]{\LARGE #1};
		\end{tikzpicture}
		\caption{#2}
		\label{#3}
	\end{figure}
}

\newcommand{\wDummy}[3]
{
	\begin{figure*}[b!]  
		\begin{tikzpicture}
		\node[draw,minimum height=6cm,minimum width=\textwidth]{\LARGE #1};
		\end{tikzpicture}
		\caption{#2}
		\label{#3}
	\end{figure*}
}

\newcommand{\basicFig}[7]
{
	\begin{figure}[#1]  	
		\vspace{#6}
		\centering		  
		\includegraphics[width=#7\columnwidth]{#2}
		\caption{#3}
		\label{#4}
		\vspace{#5}	
	\end{figure}
}
\newcommand{\fig}[4]{\basicFig{#1}{#2}{#3}{#4}{0cm}{0cm}{1}}
\newcommand\sFig[2]{\begin{subfigure}{#2}\includegraphics[width=\textwidth]{#1}\caption{}\end{subfigure}}

\newcommand{\subfig}[3]
{
	\subfloat[#3]
	{
		\includegraphics[width=#2\textwidth]{#1}
	}
	\hfill
}

\newcommand\circled[1] 
{
	\tikz[baseline=(char.base)]
	{
		\node[shape=circle,draw,inner sep=1pt] (char) {#1};
	}\xspace
}

%
%
\begin{acronym}
	\acro{CATWOMAN}{Coding applied to wireless on mobile ad-hoc networks}
	\acro{OGM}{Originator Message}
	\acro{ELP}{Echo Location Protocol}
	\acro{TQ}{Transmission Quality}
	\acro{EWMA}{Exponentially Weighted Moving Average}
	\acro{DDD}{Distributed Dispersion Detection}
\end{acronym}

\newcommand\catwoman{\ac{CATWOMAN}\xspace}
\newcommand\ogm{\ac{OGM}\xspace}
\newcommand\ogms{\acp{OGM}\xspace}
\newcommand\elp{\ac{ELP}\xspace}
\newcommand\tq{\ac{TQ}\xspace}
\newcommand\ewma{\ac{EWMA}\xspace}
\newcommand\ddd{\ac{DDD}\xspace}

%
%
\begin{acronym}
	
	%
	%
	\acro{KPI}{Key Performance Indicator}
	\acro{PDR}{Packet Delivery Ratio}
	
	%
	%
	\acro{LTE}{Long Term Evolution}
	\acro{mmWave}{millimeter Wave}
	\acro{MANET}{Mobile Ad-hoc Network}
	\acro{VANET}{Vehicular Ad-hoc Network}
	\acro{V2X}{Vehicle-to-Everything}
	
	%
	%
	\acro{ANN}{Artificial Neural Network} 
	\acro{SVM}{Support Vector Machine} 
	
	%
	%
	\acro{AODV}{Ad hoc On-demand Distance Vector}
	\acro{B.A.T.M.A.N.}{Better Approach To Mobile Ad-hoc Networking} 
	\acro{GPSR}{Greedy Perimeter Stateless Routing}
	\acro{OLSR}{Optimized Link State Routing}
	
	%
	%
	\acro{OMNeT++}{Objective Modular Network Testbed in C++}  
	\acro{LIMoSim}{Lightweight ICT-centric Mobility Simulation}

	%
	%
	\acro{MAC}{Medium Access Control}
	\acro{UDP}{User Datagram Protocol}
	\acro{TCP}{Transmission Control Protocol}
	\acro{CBR}{Constant Bitrate}
	
	%
	%
	\acro{UAV}{Unmanned Aerial Vehicle} 
	\acro{ITS}{Intelligent Transportation System}
	
\end{acronym}

\newcommand\kpi{\ac{KPI}\xspace}
\newcommand\pdr{\ac{PDR}\xspace}
\newcommand\lte{\ac{LTE}\xspace}
\newcommand\mmWave{\ac{mmWave}\xspace}
\newcommand\manet{\ac{MANET}\xspace}
\newcommand\vanet{\ac{VANET}\xspace}
\newcommand\manets{\acp{MANET}\xspace}
\newcommand\vanets{\acp{VANET}\xspace}
\newcommand\ann{\ac{ANN}\xspace}
\newcommand\svm{\ac{SVM}\xspace}
\newcommand\aodv{\ac{AODV}\xspace}
\newcommand\batman{\ac{B.A.T.M.A.N.}\xspace}
\newcommand\gpsr{\ac{GPSR}\xspace}
\newcommand\olsr{\ac{OLSR}\xspace}
\newcommand\omnet{\ac{OMNeT++}\xspace}
\newcommand\inet{\texttt{INET}\xspace}
\newcommand\inetmanet{\texttt{INETMANET}\xspace}
\newcommand\limosim{\ac{LIMoSim}\xspace}
\newcommand\mac{\ac{MAC}\xspace}
\newcommand\udp{\ac{UDP}\xspace}
\newcommand\tcp{\ac{TCP}\xspace}
\newcommand\cbr{\ac{CBR}\xspace}
\newcommand\uav{\ac{UAV}\xspace}
\newcommand\uavs{\acp{UAV}\xspace}
\newcommand\vtox{\ac{V2X}\xspace}
\newcommand\its{\ac{ITS}\xspace}

\acresetall
\title{\paperTitle}

\author{\IEEEauthorblockN{\textbf{\paperAuthors}}
	\IEEEauthorblockA{Communication Networks Institute,	TU Dortmund University, 44227 Dortmund, Germany\\
		e-mail: \paperEmails}}

\maketitle

%
%
\def\COPYRIGHTYEAR{2019}
\def\CONFERENCE{2019 IEEE 89th IEEE Vehicular Technology Conference (VTC-Spring)} 
\def\DOI{10.1109/VTCSpring.2019.8746361}	
\def\bibtex
{
	@InProceedings\{Sliwa/etal/2019a,
	author    = \{Benjamin Sliwa and Stefan Falten and Christian Wietfeld\},
	title     = \{Performance evaluation and optimization of \{B.A.T.M.A.N. V\} routing for aerial and ground-based mobile ad-hoc networks\},
	booktitle = \{2019 IEEE 89th Vehicular Technology Conference (VTC-Spring)\},
	year      = \{2019\},
	address   = \{Kuala Lumpur, Malaysia\},
	month     = \{April\},
	\}
}
\ifx\CONFERENCE\VOID
\def\conferencenotice{Submitted for publication}
\def\copyrightnotice{}
\else
\ifx\DOI\VOID
\def\conferencenotice{Accepted for presentation in: \CONFERENCE}	
\else
\def\conferencenotice{Published in: \CONFERENCE\\DOI: \href{http://dx.doi.org/\DOI}{\DOI}

}
\fi
\def\copyrightnotice{
	\copyright~\COPYRIGHTYEAR~IEEE. Personal use of this material is permitted. Permission from IEEE must be obtained for all other uses, including reprinting/republishing this material for advertising or promotional purposes, collecting new collected works for resale or redistribution to servers or lists, or reuse of any copyrighted component of this work in other works.
}
\fi
\def\overlayimage{%
	\begin{tikzpicture}[remember picture, overlay]
	\node[below=5mm of current page.north, text width=20cm,font=\sffamily\footnotesize,align=center] {\conferencenotice \vspace{0.3cm} \pdfcomment[color=yellow,icon=Note]{\bibtex}};
	\node[above=5mm of current page.south, text width=15cm,font=\sffamily\footnotesize] {\copyrightnotice};
	\end{tikzpicture}%
}
\overlayimage
\begin{abstract}
		
%
%
The provision of reliable and efficient end-to-end communication within ground- and air-based mobile mesh networks is a major challenge for routing protocols due to the mobility-related dynamics of the channel properties and the resulting mesh network topology.
%
%
In this paper, we evaluate the performance of the novel \batman V routing protocol for vehicular mesh networks and propose a mobility-predictive extension that explicitly addresses highly dynamic communication networks. In order to enable large-scale simulative analysis, we present an open source simulation model, which is validated by field experiments.
%
%
Within a comprehensive evaluation campaign in \vtox and \uav scenarios, it is shown that 
the predictive \batman V-based approach is significantly better suited for maintaining reliable connectivity within highly mobile mesh networks than established routing protocols.

\end{abstract}

\IEEEpeerreviewmaketitle

\section{Introduction}

%
%
The usage of multi-hop communication within mesh-based \manets enables significant range extensions as well as path redundancy, which is exploited in vehicular platooning, remote sensing of disaster areas using \uavs and ad-hoc network provisioning (c.f. Fig.~\ref{fig:scenario}). Due to the potentially high relative mobility of the different nodes and the short lifetime of routing paths, which results in a highly dynamic network topology, routing is immanently challenging in these networks. 

%
%
A promising solution approach that explicitly addresses these challenges, is the usage of \emph{anticipatory communication} \cite{Bui/etal/2017a, Sliwa/etal/2018b}, which aims to utilize the existing communication network in a more efficient way by an increased situation-awareness. Cross-layer principles allow the integration of the mobility characteristics of the vehicles themselves into the routing decisions.

%
%
In this paper, we extend previous work on mobility-aware vehicular mesh routing \cite{Sliwa/etal/2016b} based on the \batman routing protocol \cite{Johnson/etal/2008a} in order to leverage new features of the fifth protocol version.
%
%
Unlike most established routing protocols, which handle routing on the network layer, \batman operates on the \mac layer since version IV \cite{Hardes/etal/2017a}. Moreover, version V introduces major changes and novelties in the routing behavior and its metrics.
%
%
\fig{b}{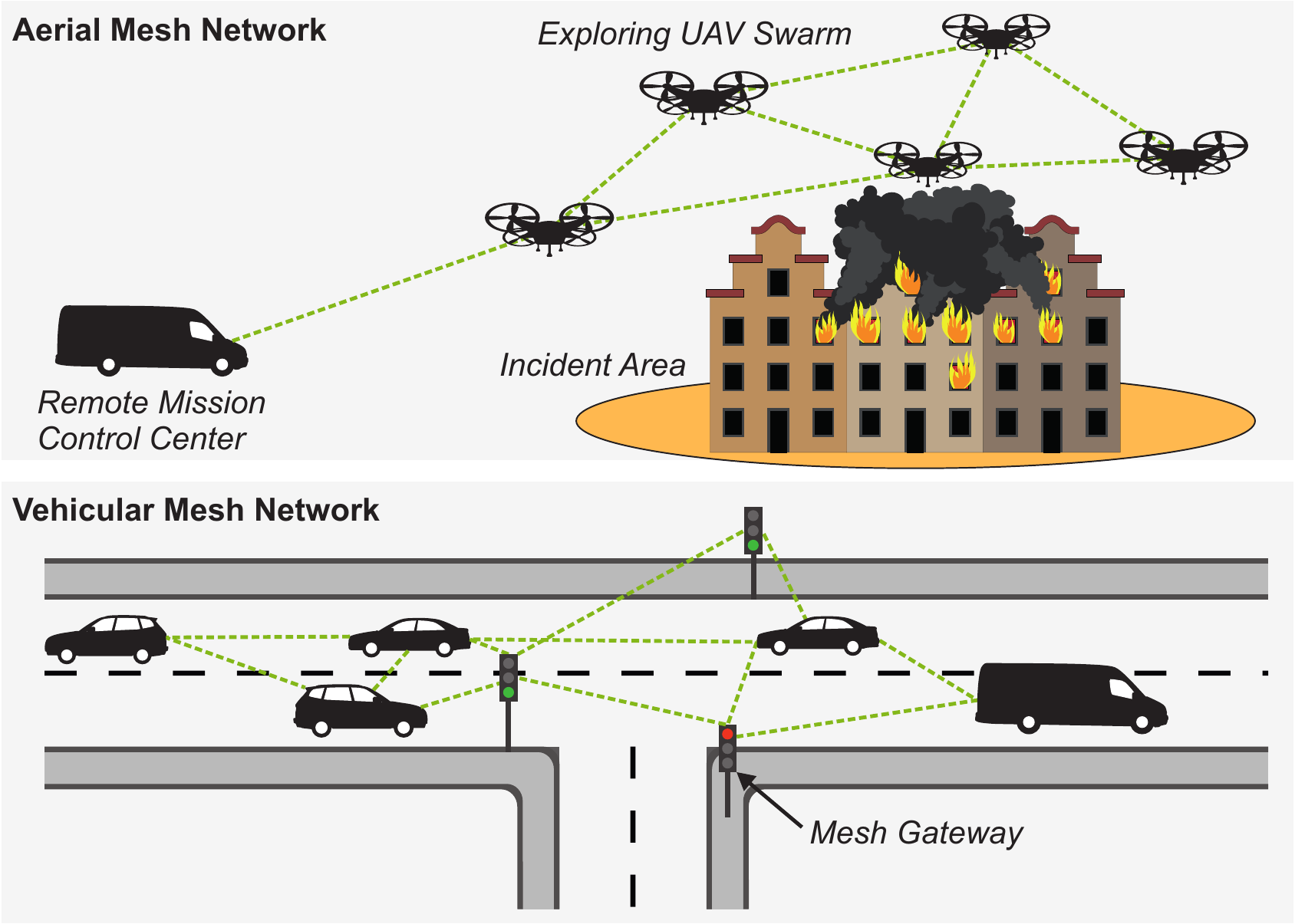}{Example applications for aerial and vehicular mesh networks.}{fig:scenario}
In the remainder of this paper, the following contributions to anticipatory mesh networking are provided:
%
%
\begin{itemize}
	\item Presentation of a novel \textbf{open source simulation model of \batman V} for \omnet \cite{Varga/Hornig/2008a} and its \inet / \inetmanet framework, which is close to the Linux kernel implementation of the protocol.
	\item Simulation model validation using \textbf{field measurements} and empirical channel modeling in Sec.~\ref{sec:validation}.
	\item Optimization and extension of \batman V for high mobility scenarios using \textbf{mobility prediction} in Sec.~\ref{sec:approach}.
	\item Performance comparison of \batman V and the proposed experimental extensions with established protocols in aerial and ground-based simulation scenarios in Sec.~\ref{sec:results}.
	\item All raw results are provided in an \textbf{open access} way.
\end{itemize}

\section{Related Work}

In this section, we discuss current approaches for vehicular mesh routing and provide information about the general properties of the considered \batman protocol.

\subsection{Overview of Current Vehicular Mesh Routing Approaches} \label{sec:state_of_the_art}

%
%
A detailed overview about current \vanet -- which form a subclass of \manets -- trends and research topics based on literature analysis is provided by \cite{Cavalcanti/etal/2018a}. While routing is pointed out as one of the major \vanet research topics, it is stated that traditional (topology-based) routing protocols mostly cannot satisfy the requirements of highly mobile vehicular scenarios. Similarly, the authors of \cite{Awang/etal/2017a} emphasize the need for cross-layer design techniques in order to increase the situation-awareness for routing in mobile networks.
%
%
Geo-based approaches such as \gpsr \cite{Karp/Kung/2000a} exploit positioning information for routing decisions, which has been demonstrated to work efficiently in vehicular networks. Consequently, predictive extensions have been proposed, which go another step further by integrating locations forecasts into the routing process for estimating link stability times \cite{Singal2017, Yang/etal/2017a}. Those approaches are usually implemented as an extension to existing protocols -- e.g., based on \aodv \cite{Perkins/Royer/1999a} in \cite{Hu/etal/2009a} and \gpsr-based in  \cite{Huang/Zhang/2013a}. However, although existing solutions achieve significant improvements in the resulting transmission efficiency,  the approaches are mostly generic and rely on naive predictions, e.g., extrapolations based on position, direction and velocity.

%
%
Although in the past the research in vehicular and aerial mesh networks has mostly been carried out with an isolated view on the intended usecase, recent works emphasize their similarities \cite{Menouar/etal/2017a} and advocate for a joint consideration of these vehicle types within upcoming \its scenarios, e.g., near-field delivery and ad-hoc network provisioning and sensing at road incidents by \uavs.

\subsection{General Properties of the \batman Routing Protocol}

%
%
\batman \cite{Perkins/Royer/1999a} has been proposed by the \emph{Freifunk} community after the application of \olsr \cite{Clausen/Jacquet/2003a} did not fulfill the performance requirements of large-scale mesh deployments. Although \batman is a proactive protocol and shows a similar performance as other established protocols, it has some unique properties that arise from its \emph{bio-inspired} nature, which implement the idea of \emph{stigmergy}.
\begin{itemize}
	\item \batman strictly follows a decentralized routing approach: Instead of requiring complete knowledge about the network topology, only information about the suitability of the direct neighbors for reaching a defined destination is maintained, the network in between is considered a black box. As information is redundantly received from all other nodes of the network -- which are referred to as \emph{originators} -- via multiple paths through the immediate neighboring nodes, it is highly robust against packet loss.	
	\item The periodically transmitted \ogms, which contain information about the \emph{reverse path quality}, are propagated through the whole network. The metric calculation and message handling can easily be exchanged by another method, making it a promising research platform for optimization and development of new link quality assessment methods.
	\item Since multiple paths are maintained for each destination, multi-path communication can be integrated for load balancing \cite{Sliwa/etal/2017a}.
	\item In contrast to other protocols, which are often only evaluated in simulators, the development of \batman V is driven by a practical usecase intention and implementations are available for Linux-based operating systems.
\end{itemize}

\section{Simulation-based System Model} \label{sec:approach}

In this section, the simulation model and its extensions for highly mobile scenarios are presented. 

%
%

%
%
\fig{}{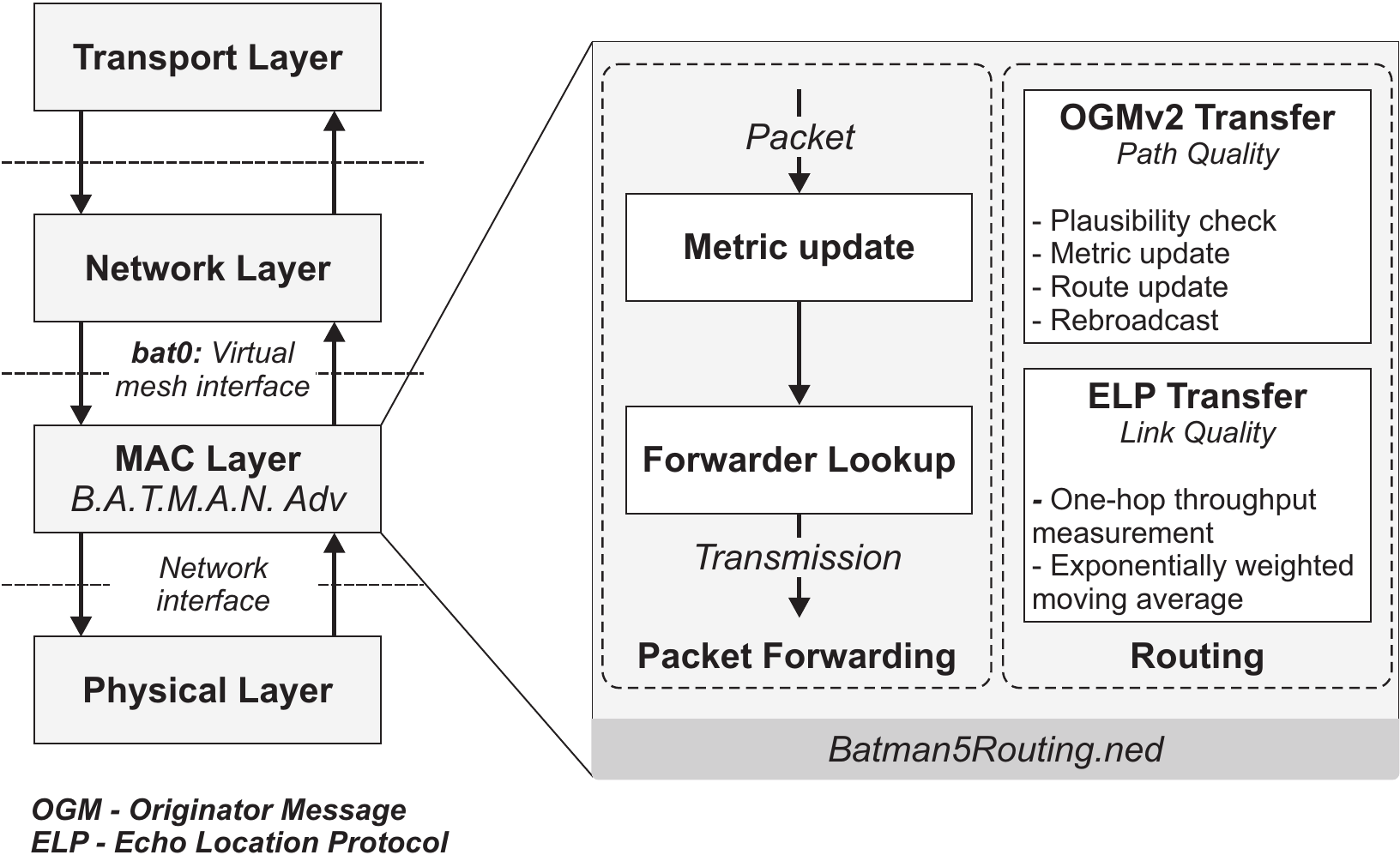}{Architecture model and message flow for \batman V routing.}{fig:approach}
%
%
An illustration of the overall architecture model and the message flow is shown in Fig.~\ref{fig:approach}.
%
%
As an entry point for all exchanged packets, the mesh interface \emph{bat0} is used to abstract the underlying mesh topology of the network, acting as a virtual network switch. All routing messages are encapsulated in raw Ethernet frames.
%
%
The \mac layer approach allows the implementation of the routing mechanism as a \emph{linux kernel module}, which -- in contrast to network layer routing -- avoids the necessity for packet exchange between userspace and kernelspace, reducing computing effort and energy consumption.

%
%
Although the \texttt{INETMANET} framework of \omnet provides \texttt{ManetRoutingBase.ned} as a common abstraction for \manet protocols, it only considers network layer routing. Therefore, the simulation model is based on a novel logical host module \texttt{Batman5Router.ned}, which links the interfaces of the different logical layers.

The routing protocol uses different packet types to determine the \emph{reverse path metric} to the originators within the network, which is further explained in the following paragraphs.

\subsection{Link Quality Assessment}

Due to the decentralized approach of \batman V, all decision processes are based on link-level knowledge, which is obtained by periodical exchange of \elp messages. In regular protocol implementation, the latter are used to estimate the throughput to the direct neighbors. Additionally, \emph{\elp probing} triggers the transmission of two unicast \elp messages to each forwarder within the defined \elp interval in order to artificially generate traffic for the throughput estimation. While this procedure is mandatory in the linux kernel implementation, it is optional in the simulation model and can be configured to use variable packet sizes, which are further evaluated in Sec.~\ref{sec:results}. As an alternative to the throughput metric, a hopcount-based approach, which uses hop penalties, can be applied.
After node $N$ has determined new immediate link metric value $\Phi^{N}_{F}$ to forwarder $F$, the final value is calculated based on the link quality history using \ewma.

\subsection{Routing Process and Metrics}

%
%
Each node $N$ maintains a \emph{link metric} $\Phi^{N}_{F}$ for each of its 1-hop neighbors, based on knowledge obtained from \elp message exchange. 
%
%
Routing decisions are based on the end-to-end \emph{path metric} towards each originator, which is derived from the metrics of the \emph{reverse path} obtained from the \ogms. Upon creation of an \ogm, each originator initializes the reverse path score as the \texttt{uint32} maximum $2^{32}-1$ and announces it with the message. The intermediate nodes update the contained value based on their local knowledge about the link quality to the last forwarder. In the proposed simulation model, the metric calculation is formulated in an abstract way in order to emphasize its \emph{platform} character, which enables the integration of further metrics in the future.  As all logical calculations are performed in the continuous domain in the range $\left[ 0, 1\right]$, the \texttt{uint32} values are transformed into this value range for the processing step and back before the packet forwarding.

Upon reception of an \ogm of node $D$ via a forwarder $F$, node $N$ computes the reverse path metric $\Psi^{N}_{D|F}$ as
%
%
\begin{equation} \label{eq:path_metric}
	\Psi^{N}_{D|F} = \Theta(\hat{\Psi}_{D}, \Phi^{N}_{F})
\end{equation}
by applying an operator $\Theta$ to the received reverse path metric $\hat{\Psi}_{D}$ and the link metric $\Phi^{N}_{F}$ to the forwarder. A summary of the considered metrics for the different \batman protocol versions and the proposed extensions is given in Tab.~\ref{tab:metrics}.
%
%
\newcommand{\entry}[2]{#1 & #2\\}
\newcommand{\head}[2]{\toprule \entry{\textbf{#1}}{\textbf{#2}}\midrule}

\newcommand\TW[1]{#1\columnwidth}

\begin{table}[ht]
	\centering
	\caption{Established and experimental routing metrics for the different \batman versions}
	\begin{tabular}{p{\TW{0.1}}p{\TW{0.7}}}
		
		\head{Version}{Metric}
		
		\entry{III}{Average number of received \ogms within a defined interval}
		\entry{IV}{Determination of the \tq}
		\entry{V}{Throughput estimation based on \elp transfer}
		\entry{V}{Hop count}
		
		\midrule
		
		\entry{V}{Distance-based forwarding using position information}
		\entry{V}{Mobility prediction}

		\bottomrule
		
	\end{tabular}
	\label{tab:metrics}
\end{table}

\subsubsection{Distance-based Forwarding}

As discussed in \ref{sec:state_of_the_art}, the usage of geo-information is a promising approach for routing in vehicular networks. Therefore, the basic \batman V protocol is extended for geo-based decision making through a novel routing metric, which is based on position announcements within the \elp messages.

With $d^{N}_{F}(t)$ being the distance from node $N$ to node $F$ and $d_{\text{max}}$ being an estimation for the transmission range, the proposed metric update procedure is handled as
%
%
\begin{equation} \label{eq:geo_forwarding}
	\Psi^{N}_{D|F}(t) = 
	\hat{\Psi}_{D} - 
	\left( \frac{d^{N}_{F}(t) }{d_{\text{max}}}\right)^{\alpha}
\end{equation}
with $\alpha$ as a weighting exponent.

\subsubsection{Mobility-predictive Forwarding}

%
%
In order to achieve a higher level of awareness about the future network topology, the geo-based forwarding scheme is extended by a predictive component. Upon transmission of an \elp message, each node predicts its own future position based on trajectory knowledge and announces it within the message.

%
%
In the following, it is assumed that each vehicle is able to determine an estimation of its near future trajectory. For completeness, we refer to the analysis in \cite{Sliwa/etal/2018a}, where the accuracy of different mobility prediction approaches is compared in real world vehicular scenarios. The future position $\tilde{\vec{P}}(t+\tau)$ is then obtained by virtually moving the vehicle along the path of its trajectory for the duration of $\tau$, which is further described in \cite{Sliwa/etal/2018a}.
The mobility-predictive metric is then calculated as
%
%
\begin{equation} \label{eq:geo_prediction}
	\Psi^{N}_{D|F}(t) = 
	\hat{\Psi}_{D} - 
	\max \left[ 
	\left( \frac{d^{N}_{F}(t) }{d_{\text{max}}}\right)^{\alpha},
	\left( \frac{\tilde{d}^{N}_{F}(t+\tau) }{d_{\text{max}}}\right)^{\alpha}
	\right] 
\end{equation}
with $\tau$ being the desired prediction lookahead and $\tilde{d}^{N}_{F}(t+\tau)$ being the anticipated distance between $N$ and $F$.

\section{Methodology} \label{sec:methodology}

In this section, the setups for the simulative evaluation and the experimental validation are presented.

\subsection{Evaluation Scenario}
The simulation model is based on \omnet and the \texttt{INET/INETMANET} framework and is provided in an open source way at \cite{Sliwa/2018a} alongside with the raw results of the performed evaluations. For the field test validation, common of-the-shelf embedded computers are used that run a linux-based operating system and execute the \batman V (version 2018.0) routing mechanism.
%
%
\renewcommand{\entry}[2]{#1 & #2\\}
\renewcommand{\head}[2]{\toprule \entry{\textbf{#1}}{\textbf{#2}}\midrule}

\begin{table}[ht]
	\centering
	\caption{Default Simulation Parameters}
	\begin{tabular}{ll}
		
		\head{Parameter}{Value}
		
		\entry{\mac}{IEEE 802.11g}

		\entry{Channel model: Rural / Urban}{$\left\lbrace \text{Friis}, \text{Nakagami~(m=2)}\right\rbrace $}
		\entry{Path loss exponent $\eta$}{2.65}
		\entry{Receiver sensitivity}{-83~dBm}
		
		\entry{Transmission power}{$\left\lbrace 10, 20\right\rbrace $~dBm}
		\entry{Carrier frequency}{2.4~GHz}
		\entry{Number of mesh nodes}{10}
		\entry{Stream data rate}{10~MBit/s}
		\entry{Simulation duration per run}{300~s}
		\entry{Simulation runs per setting}{25}
		\entry{Playground size: Generic scenario}{600~m x 600~m x 10~m}
		\entry{Playground size: UAV scenario}{500~m x 500~m x 250~m}
		\entry{Playground size: Vehicular scenario}{1500~m x 1000~m x 10~m}

		\midrule
		\entry{\batman version}{2018.0}
		\entry{\omnet / INET version}{5.2.1 / 3.6.3}
		\entry{\ogm interval}{0.33~s}
		\entry{\elp interval}{0.2~s}
		\entry{Weighting exponent $\alpha$: Rural / Urban}{$\left\lbrace 1, 2\right\rbrace $}
		\entry{Prediction lookahead $\tau$: Rural / Urban}{$\left\lbrace 3, 4\right\rbrace $~s}
		
		\bottomrule
		
	\end{tabular}
	\label{tab:parameters}
\end{table}

The default parameters for the different scenarios and routing protocols are summarized in Tab.~\ref{tab:parameters}. For the mobility-predictive metric, $d_{max}$ is calculated as the maximum distance that satisfies the receiver sensitivity based on the channel model. The \emph{Friis} model is used for \emph{rural} environments and the \emph{Nakagami} model is used to model \emph{urban} scenarios. The prediction lookahead is chosen based on the analysis of \cite{Sliwa/etal/2016b}. For the evaluations, the performance of a multi-hop \udp data \cbr stream is analyzed.

\subsection{Mobility Models}

For the simulative performance evaluation, a mixture of generic and vehicle-specific mobility models is applied. The initial performance evaluation and parameter optimization is performed using a Random Waypoint model in order to allow comparisons with other research works, however it is assumed that vehicles are aware of their future trajectory. For the vehicular scenario, a network provisioning scenario using gateway vehicles is modeled using \limosim \cite{Sliwa/etal/2017b}. 30 cars move around the campus area of the TU Dortmund University, which is illustrated in Fig.~\ref{fig:campus_map}. One randomly selected vehicle aims to establish a stream data connection to a network gateway vehicle using the mesh for packet forwarding.
%
%
\fig{}{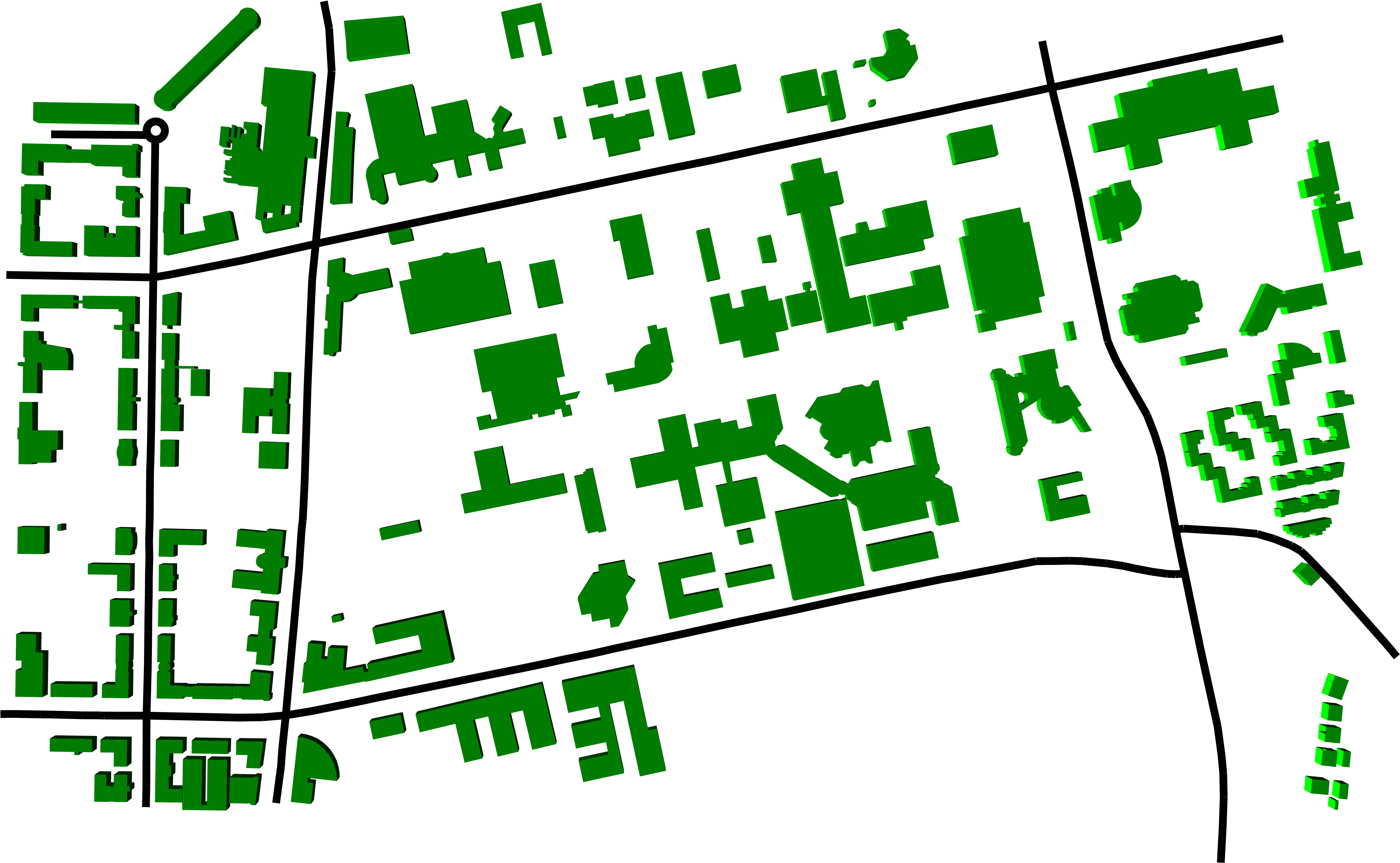}{Map of the campus area for the vehicular evaluation scenario. (Map data: ©OpenStreetMap contributors, CC BY-SA.)}{fig:campus_map}

%
%
For the evaluation of the applicability of the proposed mobility-predictive routing protocol, for low-altitude \uav applications, the \ddd \cite{Behnke/etal/2013a} algorithm applied, which is intended for agent-based plume exploration in hazardous environments. The mesh communication is used to stream video data to a remote mission control center. Here, a trace-based mobility modeling approach is applied based on traces of the \texttt{MATLAB}-based \uav simulation framework of \cite{Behnke/etal/2013a}. The \uavs make use of \emph{controlled mobility}, thus each \uav is aware of its desired trajectory in the near future.
\section{Validation} \label{sec:validation}

In this section, the derived simulation model for \batman V is validated using field measurements in controlled scenarios using the default routing metric. Within the real world measurements, a mobile vehicle transmits \udp stream data with a traffic load of 10~MBit/s to a static server node using the mesh network infrastructure. \emph{iperf} is used for traffic generation and \ac{KPI} measurement. The same scenario is modeled within a simulation setup.

%
%
\fig{}{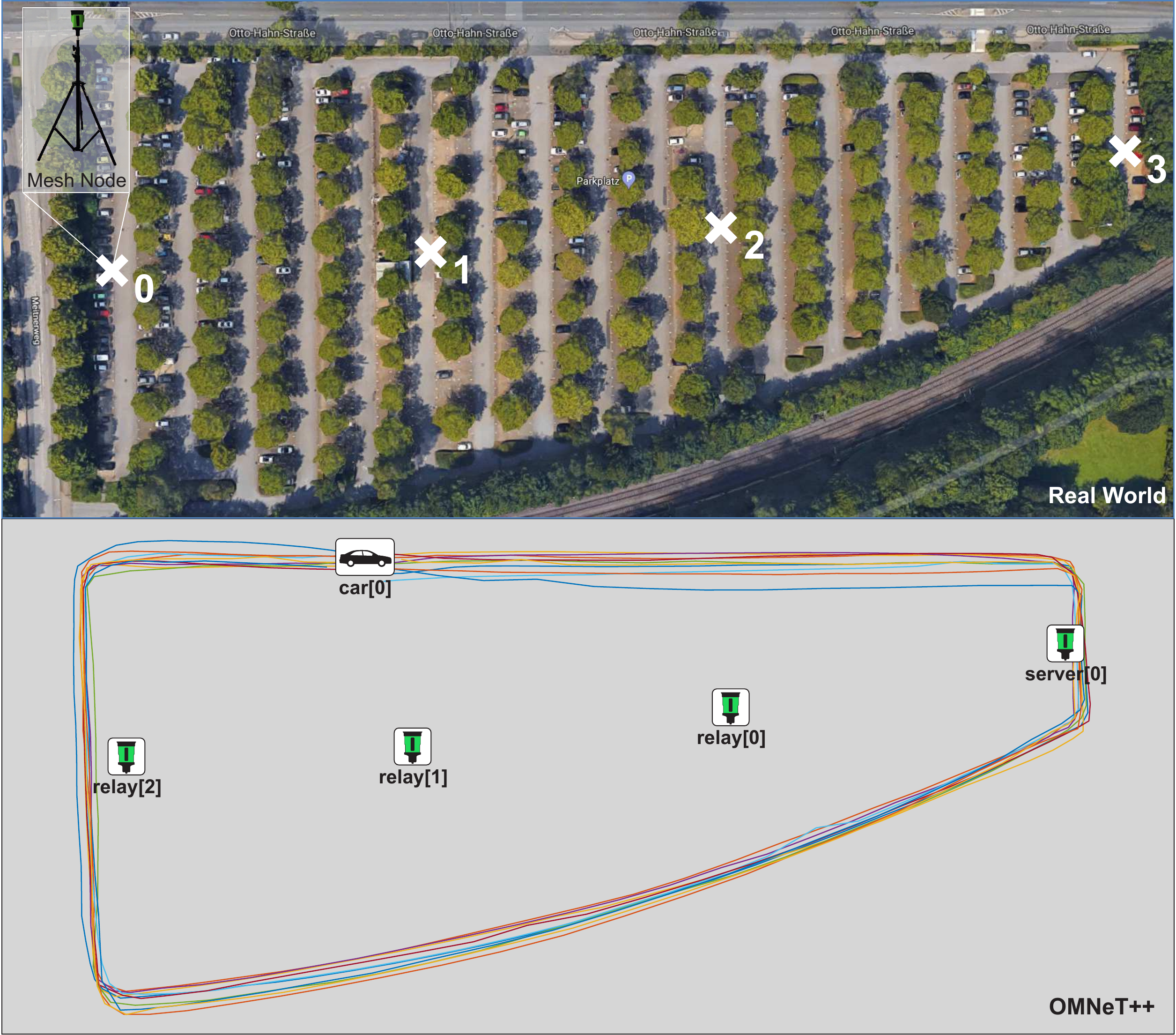}{Overview of the field test scenario and its representation within the \omnet simulator for the validation.}{fig:maps}
Fig.~\ref{fig:maps} shows a map of the real world evaluation scenario as well as its representation within the \omnet simulator. All \batman V nodes are mounted on a tripod at 1.5~m height. 
%
%
For the \emph{One-hop} scenario, the server is placed centrally between the location markers 1 and 2. \emph{Two-hop} uses location 2 for the server and location 1 for a single relay node. \emph{Multi-hop} applies the topology illustrated for the \omnet scenario as shown in Fig.~\ref{fig:maps}.
%
%
%
\fig{b}{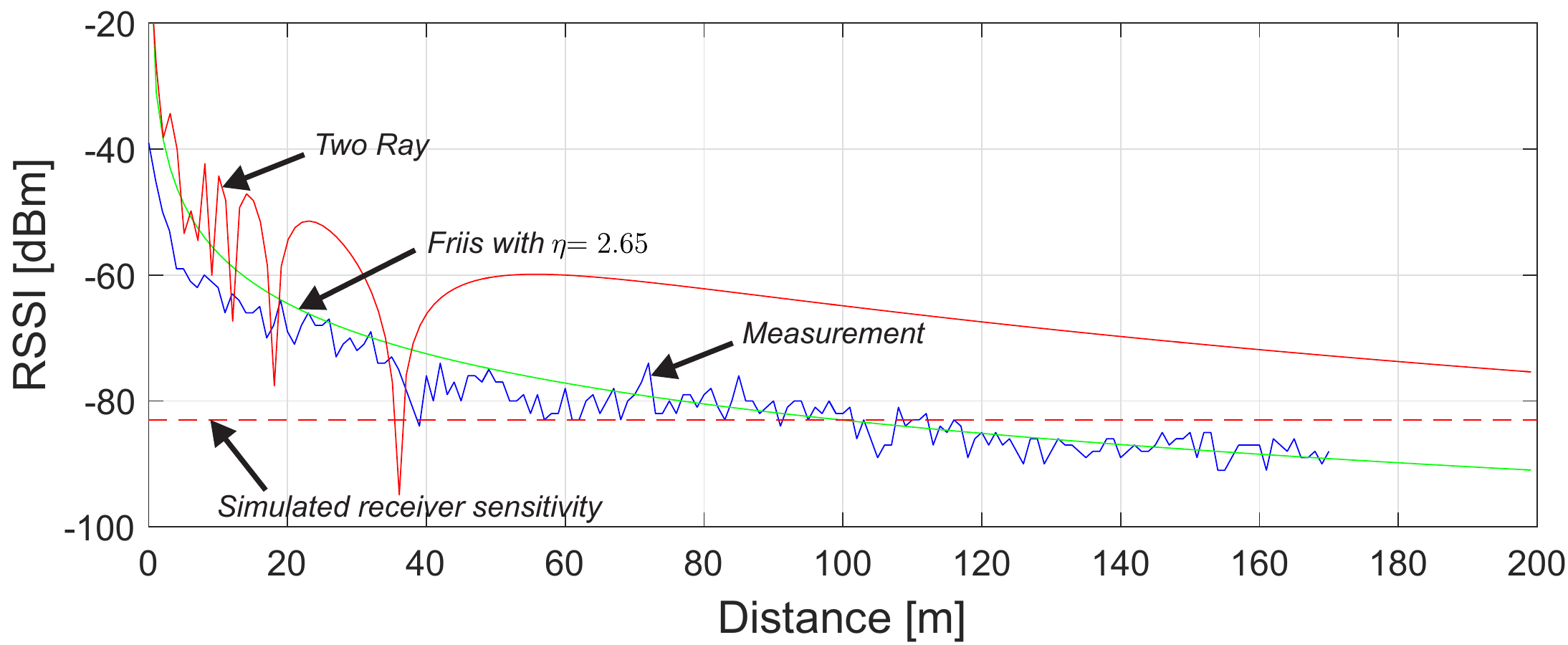}{Empirically measured channel properties and channel model approximation.}{fig:channel_model}
Fig.~\ref{fig:channel_model} shows the measured channel properties and the \emph{Friis} and \emph{Two Ray} channel models for comparison. It can be seen that the mean attenuation behavior above 70~m can be approximated by the Friis model with a path loss exponent $\eta$ of $2.65$. The Two Ray model does not fit well in the considered small-scale scenario.

%
%
In order to achieve a high congruency of the real world and the simulation scenario, an \emph{empirical channel model} is created based on the field measurements and used for the following validation.
%
%
\begin{figure*}[] 
	\centering
	\includegraphics[width=0.48\textwidth]{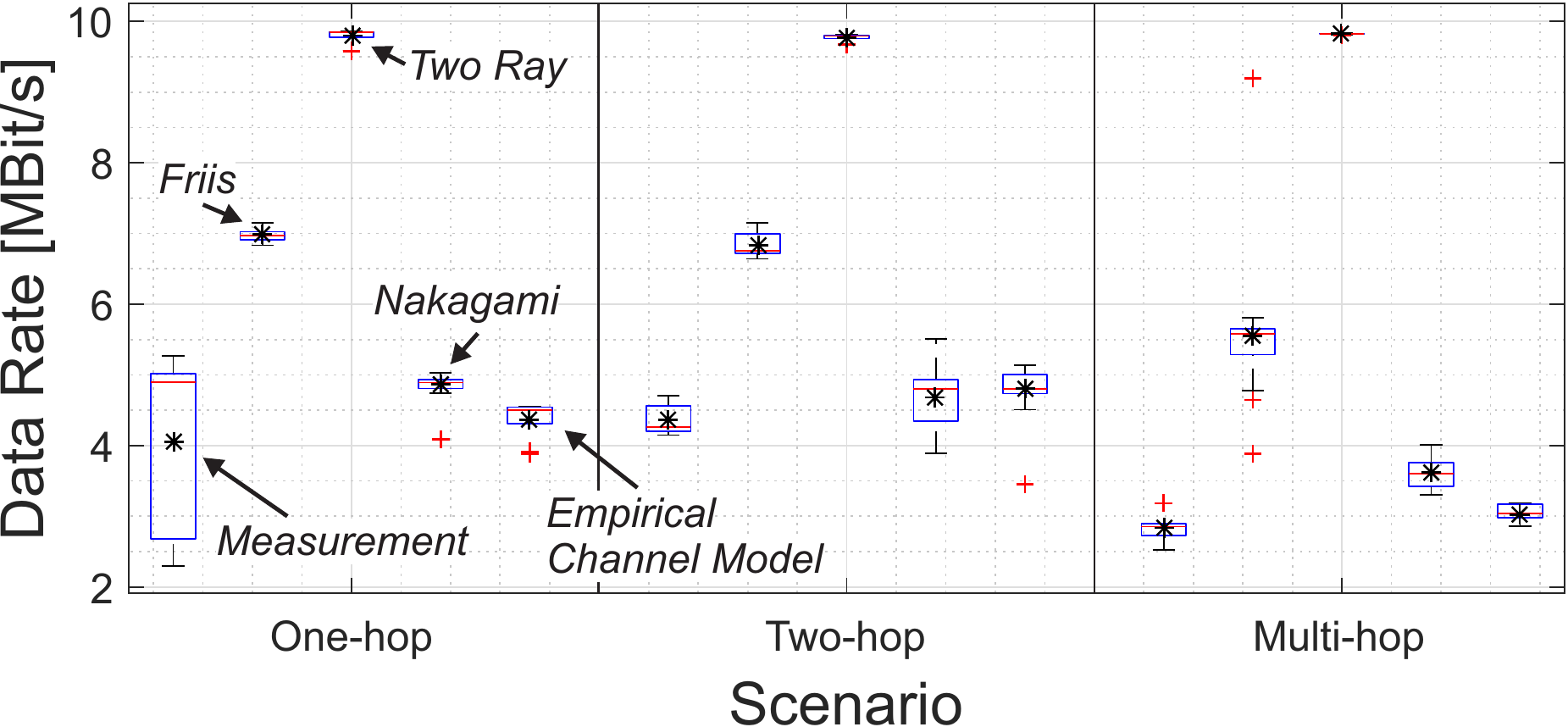}
	\includegraphics[width=0.48\textwidth]{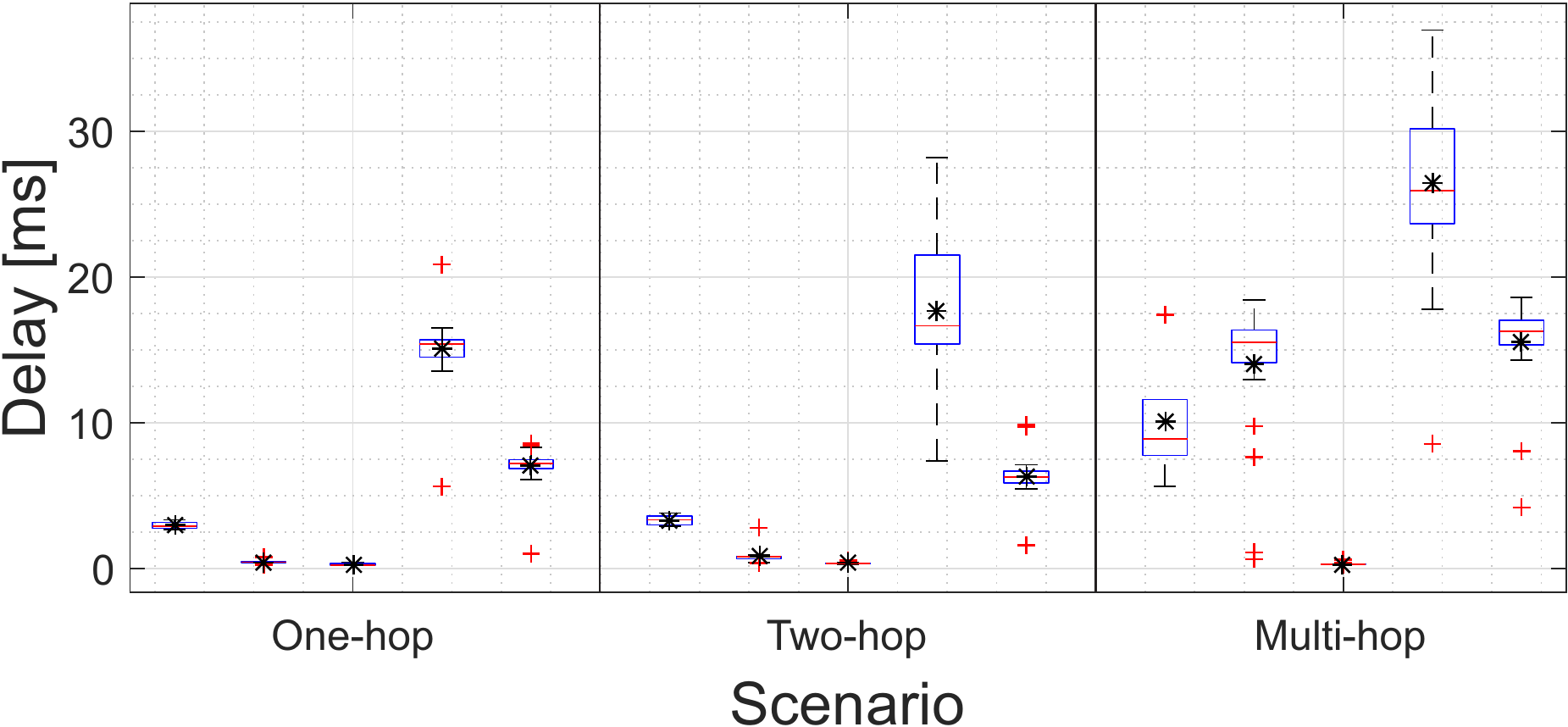}
	\caption{Model validation: Comparison of measurements for data rate and delay with simulation results using different channel models.}
	\label{fig:validation_boxes}
\end{figure*}
%
%
The resulting data rate and delay of the measurement campaign and its simulation representation is shown in Fig.~\ref{fig:validation_boxes}. The results substantiate the outcome of the channel behavior analysis. In the considered small-scale range, the Two Ray model is mostly in the \emph{constructive interference} region with an unrealistic high data rate, which is more than twice as high as the measurement, therefore it is not considered further. The Nakagami model provides a close approximation of the real world behavior of the data rate but fails to model the delay behavior with a sufficient accuracy. It can been seen that the best match for both indicators is achieved by the empirical channel model that is based on the measured channel characteristics. In conclusion, the proposed simulation model for \batman V and its real world implementation show similar average characteristics and results.

\section{Simulative Performance Evaluation} \label{sec:results}

In this section, \batman V and its proposed extensions are evaluated in different scenarios and compared with the established routing protocols \aodv \cite{Perkins/Royer/1999a}, \olsr \cite{Clausen/Jacquet/2003a}, and \batman III \cite{Johnson/etal/2008a}. All errorbars show the 0.95-confidence interval of the mean value. At first, the parametrization of \batman V is optimized for the mobile scenario based on random waypoint mobility. Afterwards, the performance of the protocol is evaluated in two reference scenarios focusing on vehicular and \uav mobility.

\subsection{Optimization for Highly Mobile Scenarios}

\batman V introduces \elp probing, which is intended to provide the throughput metric with further traffic for the link quality assessment. Therefore, the effects of the additional messages are evaluated.
%
%
\fig{}{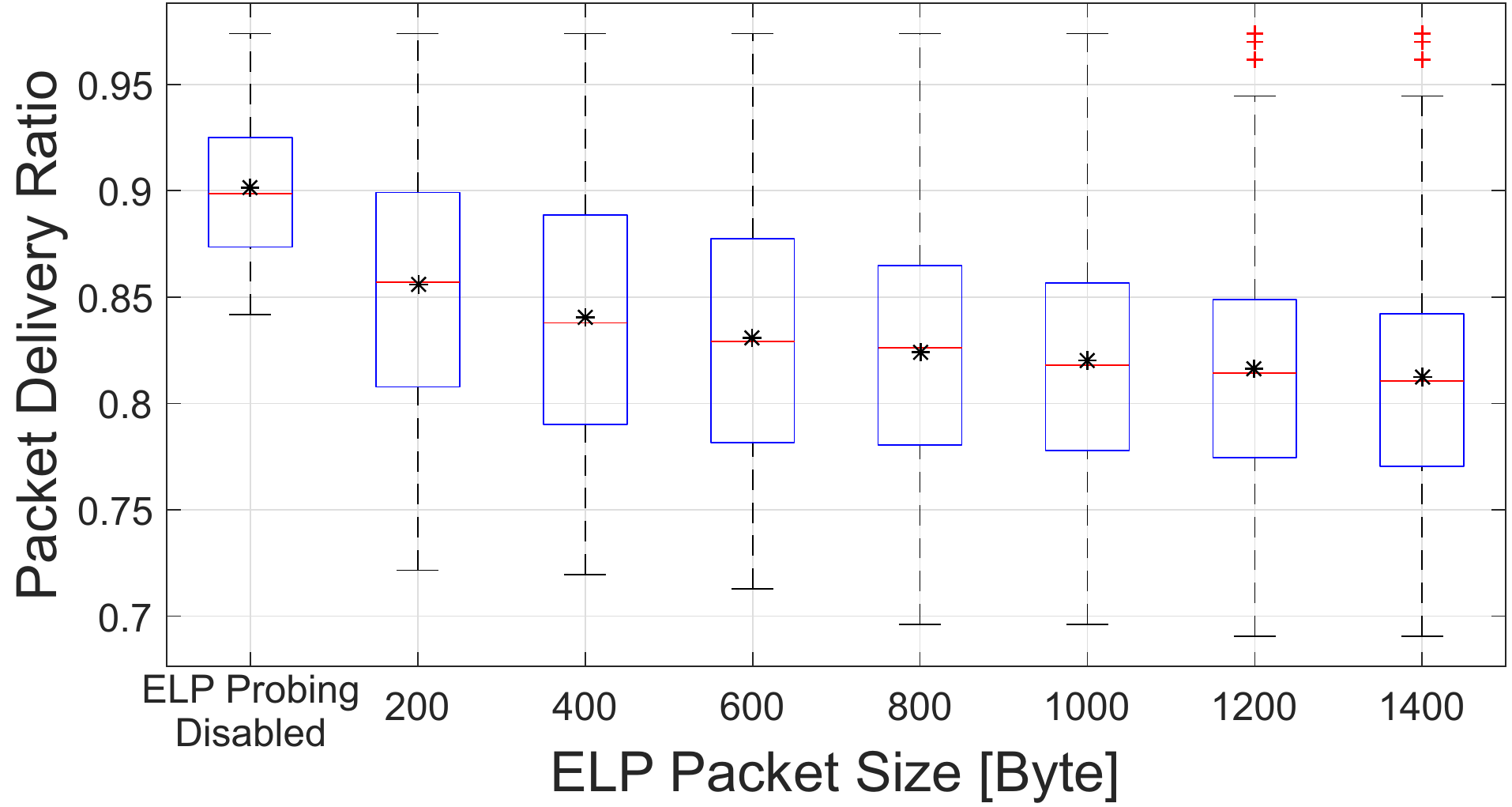}{Impact of \elp probing on the resulting data rate for different \elp packet sizes.}{fig:elp_box}
Fig.~\ref{fig:elp_box} shows the impact of \elp probing on the resulting data rate for different \elp packet sizes. It can be seen that the amount of additional messages has a negative impact on the routing performance in the mobile scenario due to the increased amount of collisions. Therefore, \elp probing is disabled for the following evaluations.

%
%
\fig{}{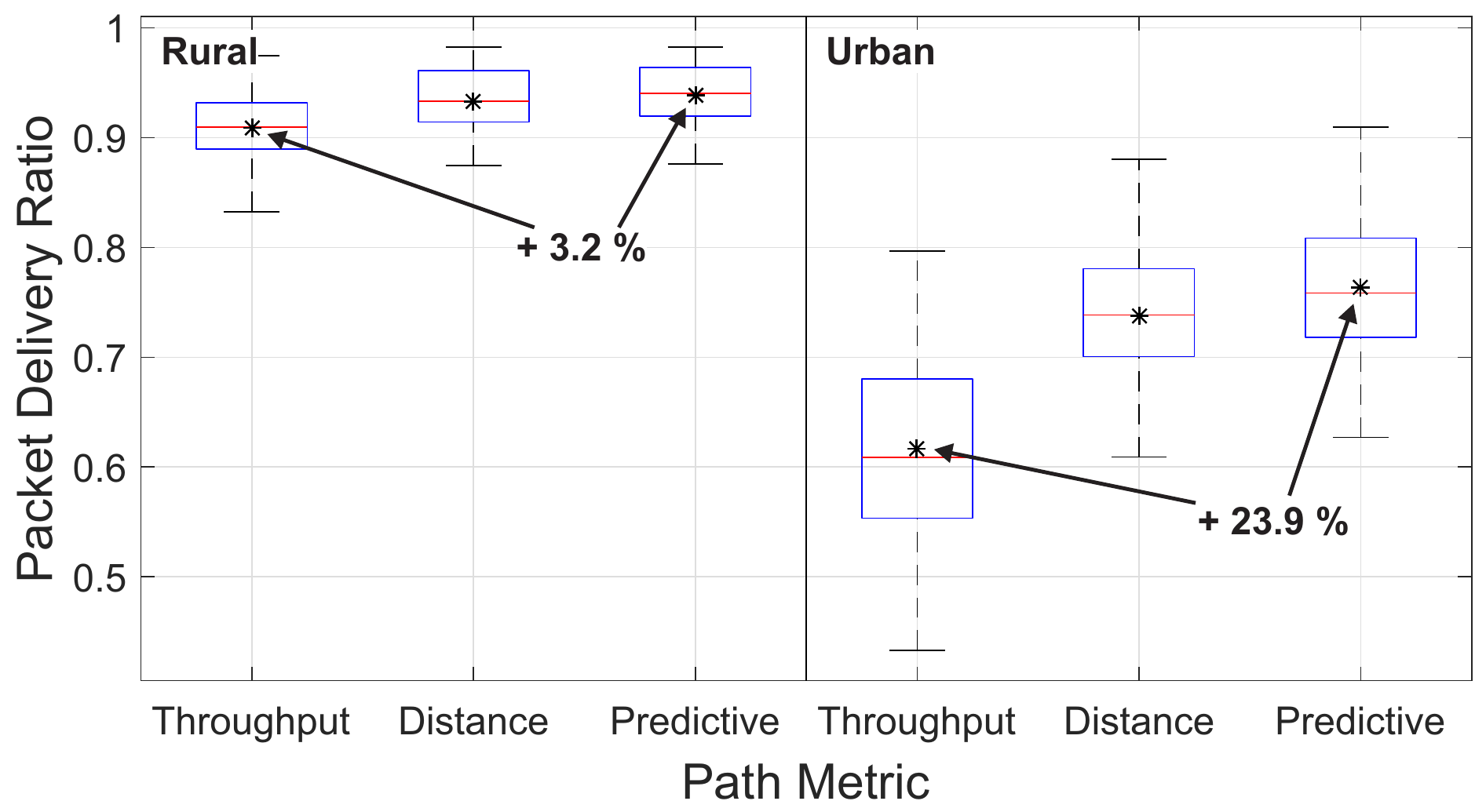}{Comparison of the native \emph{Throughput} metric with the proposed \emph{Distance} and \emph{Prediction} methods for the rural and the urban evaluation scenario.}{fig:metrics}
A comparison of the native throughput metric with the proposed mobility-aware extensions is shown in Fig.~\ref{fig:metrics}. Although both proposals outperform the native approach, the general \pdr in the rural scenario is very high, thus limiting the optimization potential. In the urban scenario, the throughput-based approach suffers from paket loss. Here, the mobility-aware approaches are able to achieve significant benefits, with the predictive approach increasing the \pdr by 23~\%. Therefore, only the proposed predictive \batman V metric will be considered in the further evaluations.

%
%
\begin{figure*}[] 
	\centering
	\includegraphics[width=\triple]{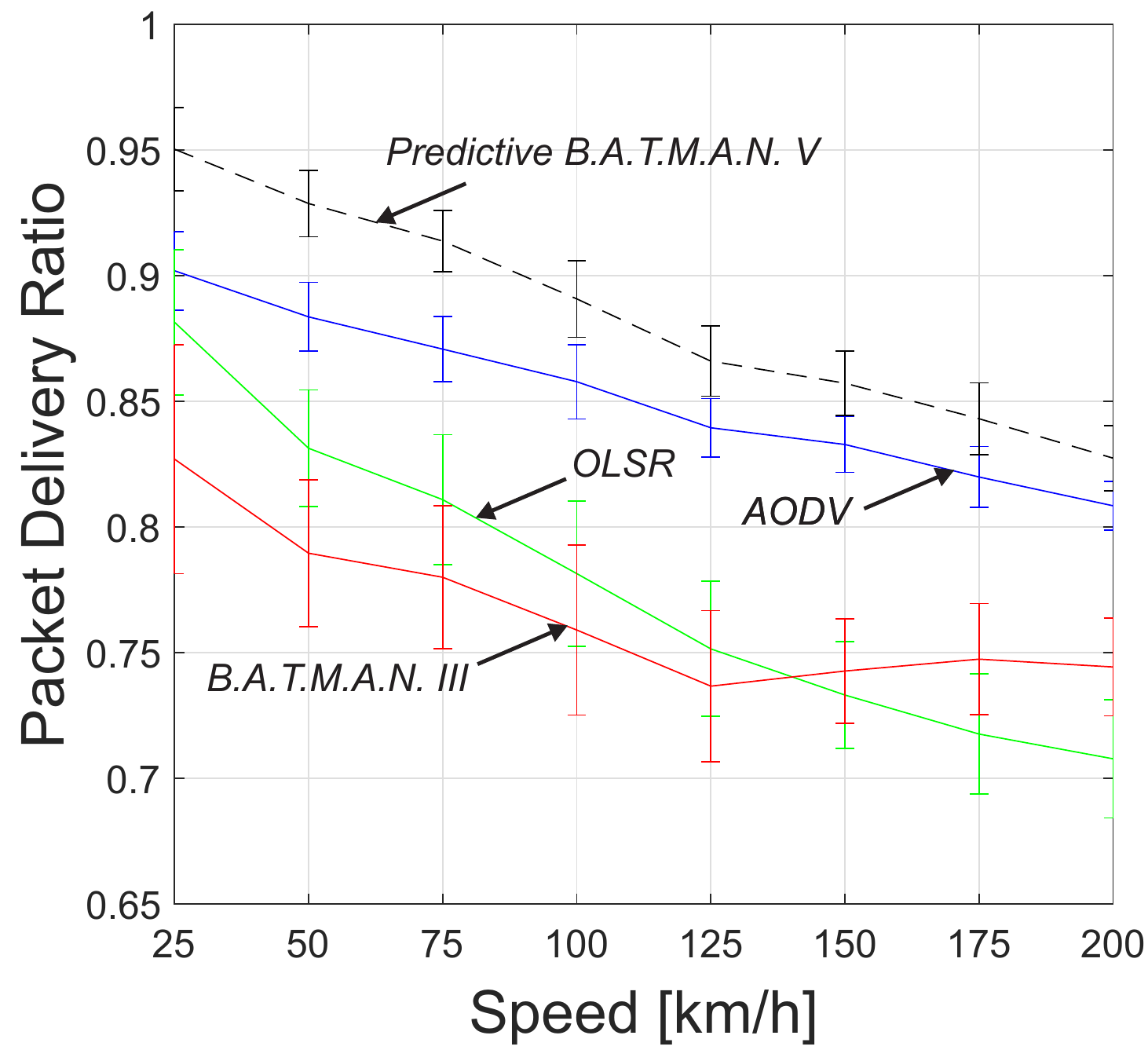}
	\includegraphics[width=\triple]{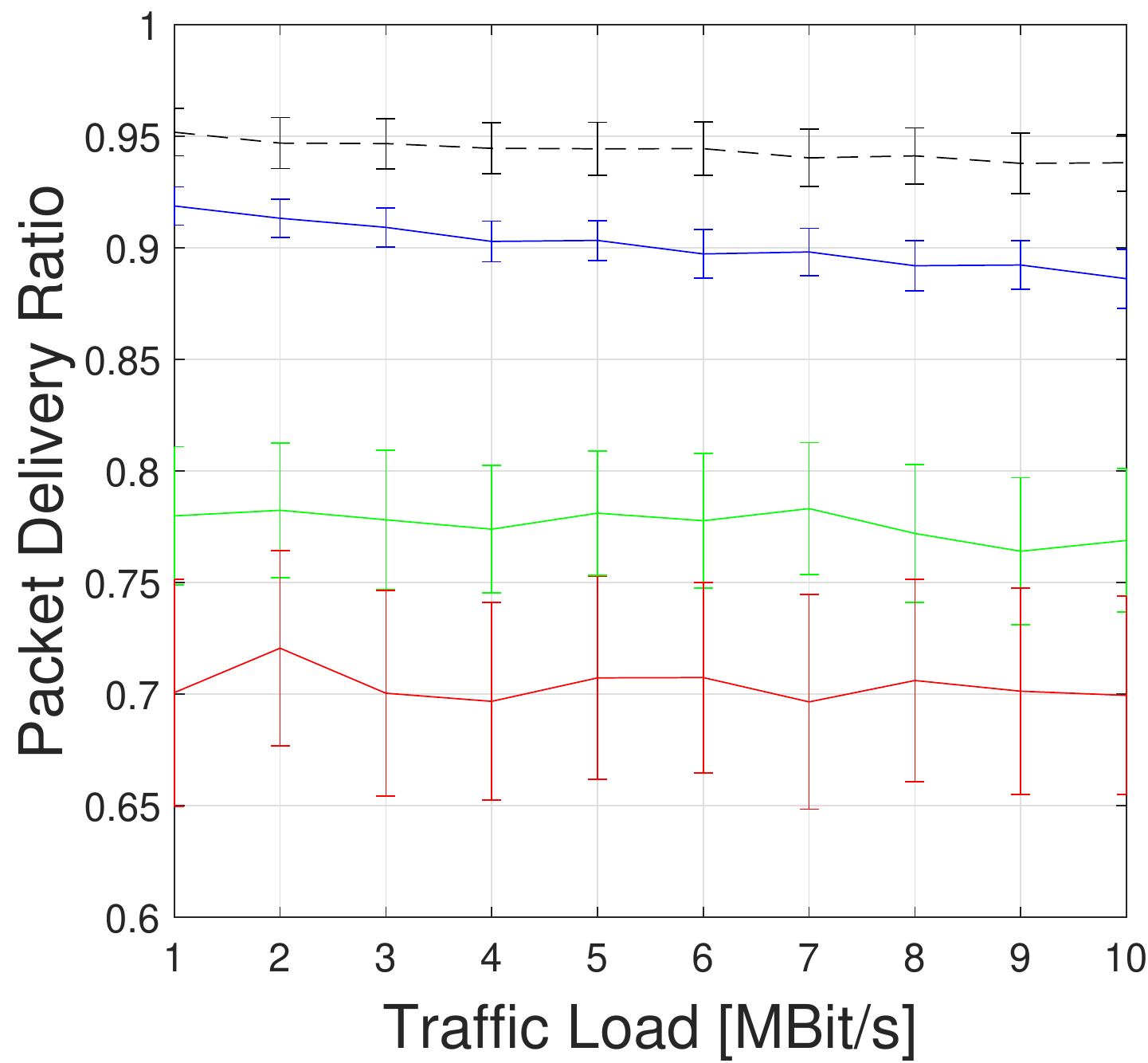}
	\includegraphics[width=\triple]{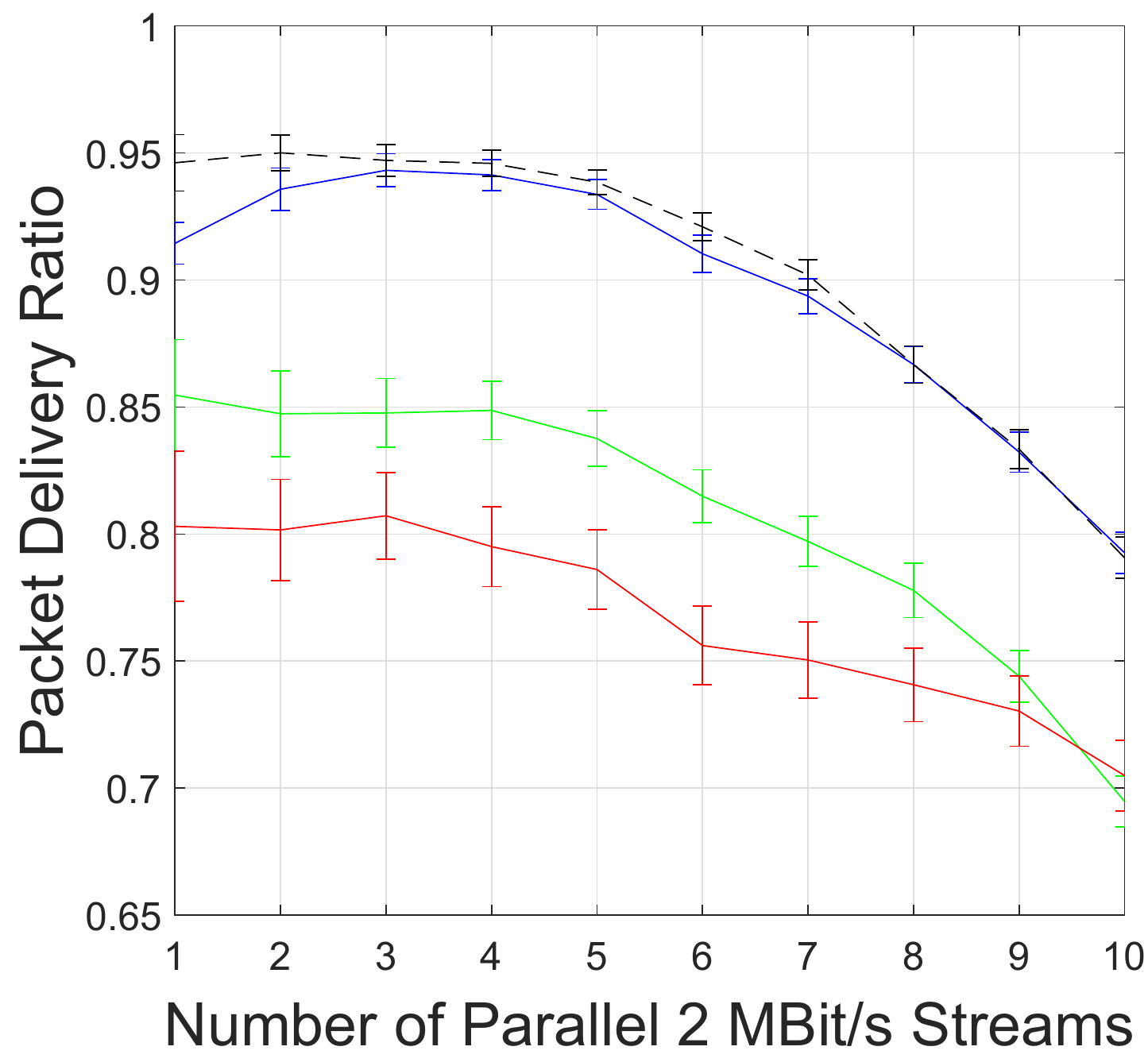}
	\caption{Scalability analysis: Comparison of the resulting \pdr with respect to vehicle speed, traffic load and number of streams for the proposed mobility-predictive \batman V and different established protocols (Rural channel).}
	\label{fig:pdr_eval}
\end{figure*}
Fig.~\ref{fig:pdr_eval} shows a scalability analysis with varied vehicle speed, traffic load and number of streams. In general, the extended \batman V and reactive \aodv achieve significantly higher \pdr values than \olsr and \batman III.
%
%
Due to its mobility-related metric, the proposed extended \batman V has a higher dependency to the velocity than the topology-based protocols, resulting in a steeper \pdr decrease with increased speed.
%
%
With the traffic load of the single stream being varied, the \pdr is only slightly reduced for all considered protocols. 
%
%
Furthermore, the behavior of the protocol for a varied number of 2~MBit/s streams with random source-destination pairs is analyzed. Here, the extended \batman V behaves slightly better than \aodv. 
%
%
Up to three parallel streams, \aodv benefits from the increased traffic on multiple end-to-end paths. Although \aodv is a reactive protocol, all intermediate nodes of an active path maintain the reverse path quality for a defined time period. Therefore, the presence of multiple active paths increases the probability that the nodes are able to switch the routing path within a changed network topology without requiring a route request phase.
For a higher number of active streams, the increasing loss probability for routing messages significantly decreases the routing performance for all considered protocols.

\subsection{\uav Evaluation Scenario}

%
%
\fig{}{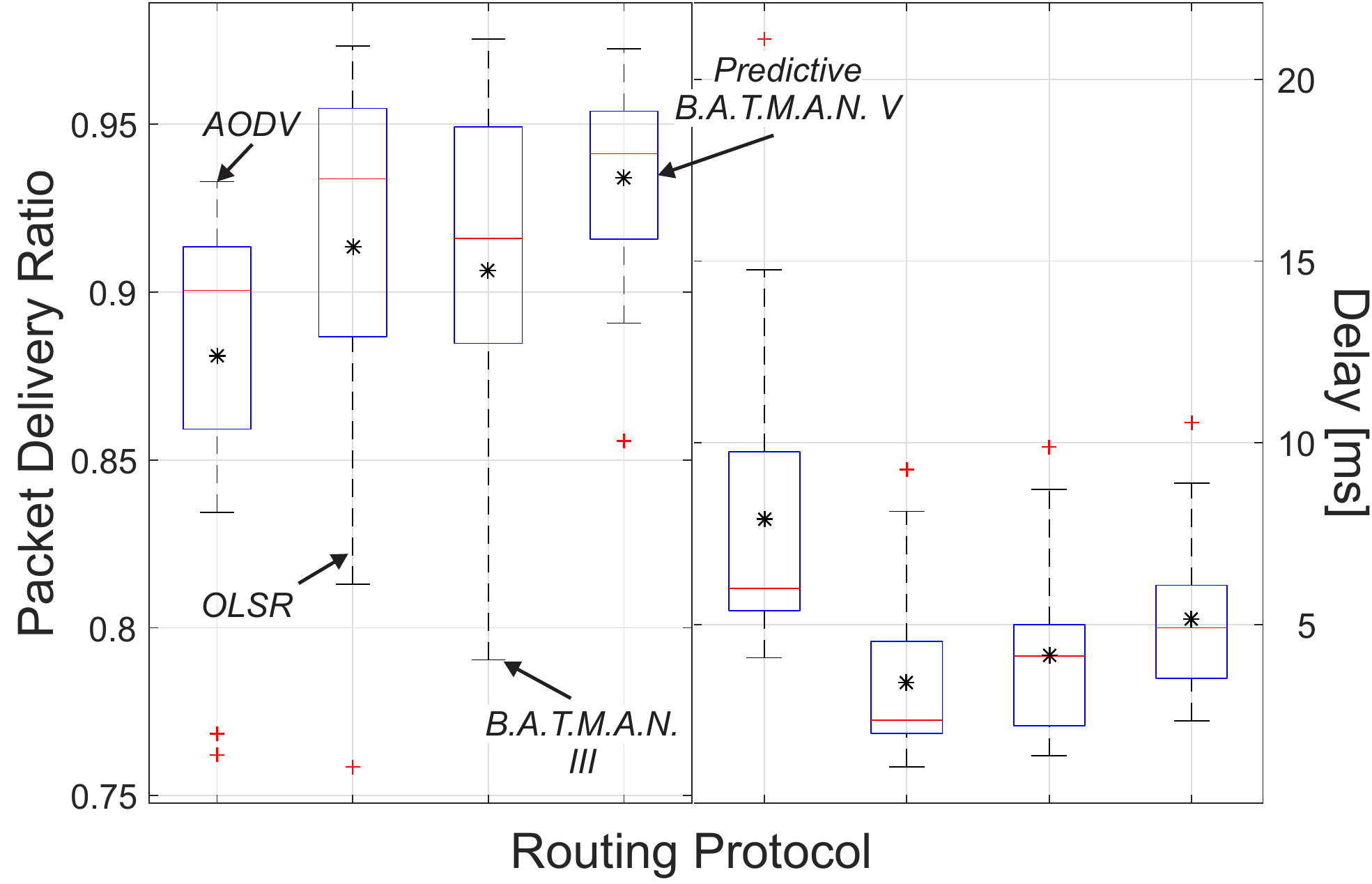}{Performance comparison using \ddd mobility for \uav-based plume exploration (Urban channel, 10 \uavs).}{fig:ddd}
Fig.~\ref{fig:ddd} shows the resulting \pdr and delay values for the considered protocols in the \uav-based plume exploration scenario using \ddd. As the mobility algorithm itself is connectivity-aware, it aims to maintain the swarm coherence by adjusting the mobility behavior of the \uavs, which results in a generally high \pdr level.
Although the mean \pdr value of the proposed approach is only slightly better than the result of \olsr and \batman III, the occurrence of lower \pdr values is significantly reduced. While all proactive protocols achieve a similar delay, the latter is significantly higher for \aodv due to its on-demand approach.

\subsection{Vehicular Evaluation Scenario}

%
%
\fig{}{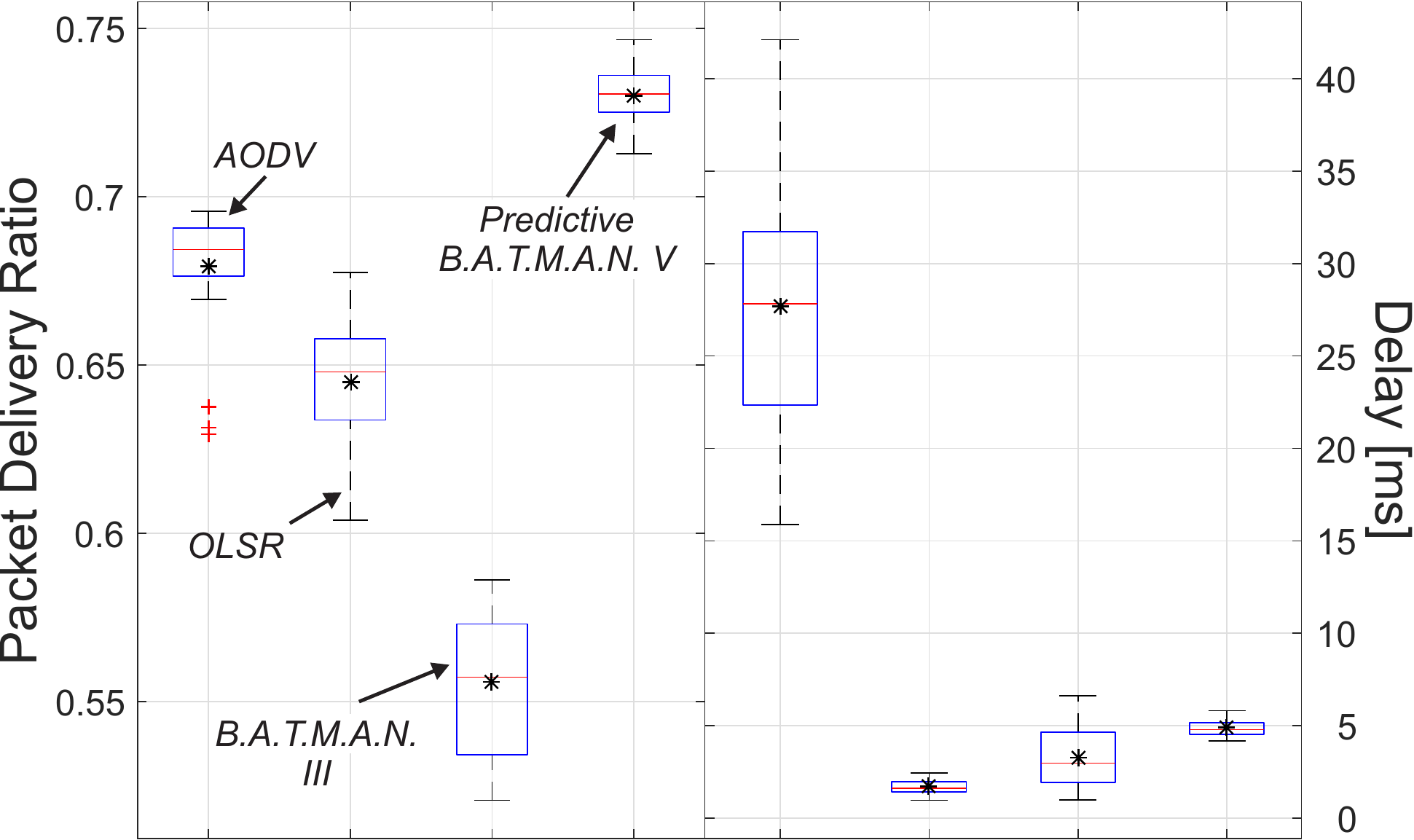}{Performance comparison for the vehicular network provisioning scenario (Urban channel, 30 vehicles).}{fig:v2x}
Fig.~\ref{fig:v2x} shows the resulting \pdr and delay values for the considered protocols.
In contrast to the previously analyzed \uav scenario, the vehicular scenario is more challenging for the routing protocols, since the larger network causes a higher number of potential routing paths from source to destination. Moreover, the physical dimensions of the scenario itself are larger and since the mobility characteristics of the vehicles are connectivity-unaware, connectivity is not always guaranteed. Therefore, the general \pdr level is lower than in the \uav scenario. 
The highest \pdr is achieved by the proposed predictive \batman V, which exploits the high predictability of road-based vehicular mobility.
Although the delay for \aodv is on average more than five times larger than for the proactive protocols, all protocols fulfill the delay ETSI requirements for safety-related vehicular communication in the considered scenario \cite{ETSI/2009a}.
\section{Conclusion}

%
%
In this paper, we presented an analysis and optimization of the novel \mac layer routing protocol \batman V for vehicular scenarios. In order to allow the simulative analysis, an open source simulation model was derived based on the Linux kernel implementation and validated using field measurements.
Within comprehensive simulation studies, it was shown that some of the new features of \batman V (e.g., \elp probing) decrease the routing efficiency in highly mobile scenarios. In addition, our proposed novel routing metrics, which exploit distance-awareness and mobility prediction, showed a significantly better performance than the throughput estimation metric of \batman V.
%
%
In different case-studies for aerial and ground-based mobile ad-hoc networks, it was shown that the extended protocol is well-suited for maintaining connectivity within vehicular and \uav-based mesh networks and outperforms established topology-based routing approaches.
%
%
In future work, we will exploit the proposed open source simulation model as a platform for optimizing vehicular mesh routing. A promising research topic is the determination of routing metrics based on reinforcement learning. Additionally, we will analyze the performance of using \batman V in scenarios with heterogeneous communication technologies.
\section*{Acknowledgment}

\footnotesize
Part of the work on this paper has been supported by Deutsche Forschungsgemeinschaft (DFG) within the Collaborative Research Center SFB 876 ``Providing Information by Resource-Constrained Analysis'', project B4.

\bibliographystyle{IEEEtran}
\bibliography{Bibliography}

\end{document}